# Hubble Space Telescope[1] Counts of Elliptical galaxies: Constraints on Cosmological Models?

Simon P. Driver[2], Rogier A. Windhorst

Department of Physics and Astronomy, Arizona State University,

Tempe, AZ 85287-1504

and

Steven Phillipps, Paul D. Bristow

Astrophysics Group, Department of Physics, University of Bristol,

Tyndall Avenue, Bristol BS8 1TL, UK

astro-ph/9511141  29 Nov 1995

---

[1]Based on observations with the NASA/ESA *Hubble Space Telescope* obtained at the Space Telescope Science Institute, which is operated by AURA, Inc., under NASA Contract NAS 5-26555.

[2]Current Address: School of Physics, University of New South Wales, Sydney, NSW 2052, Australia

## ABSTRACT

The interpretation of galaxy number counts in terms of cosmological models is fraught with difficulty due to uncertainties in the overall galaxy population (mix of morphological types, luminosity functions etc.) and in the observations (loss of low surface brightness images, image blending etc.). Many of these can be overcome if we use deep high resolution imaging of a single class of high surface brightness galaxies, whose evolution is thought to be fairly well understood. This is now possible by selecting elliptical and S0 galaxies using Hubble Space Telescope images from the Medium Deep Survey and other ultradeep WFPC2 images. In the present paper, we examine whether such data can be used to discriminate between open and closed universes, or between conventional cosmological models and those dominated by a cosmological constant. We find, based on the currently available data, that unless elliptical galaxies undergo very strong merging since $z \sim 1$ (and/or very large errors exist in the morphological classifications), then flat models dominated by a cosmological constant are ruled out. However, both an Einstein-de Sitter ($\Omega_0 = 1$) model with standard passive stellar evolution and an open ($\Omega_0 = 0.05$) model with no net evolution (*i.e.* cancelling stellar and dynamical evolution) predict virtually identical elliptical and S0 galaxy counts.

Based on these findings and the recent reportings of $H_o \simeq 80$ kms$^{-1}$ Mpc$^{-1}$, we find that the maximum acceptable age of the universe is 13.3 Gyrs and a value of $\leq 9$ Gyrs favored. A flat—$\Lambda \neq 0$—universe is therefore *not* a viable solution to the $H_o$/globular cluster age problem.

Subject headings: cosmology — galaxies: ellipticals — galaxies: evolution

## 1. Introduction

Efforts to determine cosmological parameters from galaxy number counts – effectively via measuring the volume element as a function of redshift – have been compromised by extraneous effects ever since Hubble's (1934) first attempt in this direction. Indeed, Hubble (1936) found that *no* reasonable relativistic world model appeared to fit his galaxy count data. His problem, as in many recent studies, was a combination of observational uncertainties and unknowns in the galaxy population model, in his case systematic errors in the magnitude scale and the then unknown effect of K-corrections as the ultra-violet spectrum was shifted into the blue photographic passband (see Peebles 1980).

In fact, when number counts were resurrected after a 35 year gap in the early 1970s (eg. Brown & Tinsley 1974), it was one of the "problems" which brought them back into prominence. Tinsley (1972) demonstrated that the counts were more sensitive to changes in galaxy luminosity with look-back time than to the geometry of space, and hence could be most usefully used as a probe of galaxy evolution (Tinsley 1977). Since that time, the galaxy counts have extended many magnitudes fainter (e.g., Metcalfe *et al.* 1995a), but the major obstacles to their interpretation in terms of cosmological models have remained the same. On the modelling side, these obstacles are primarily the dependence on the fraction of galaxies of different types and luminosities and the effect(s) of galaxy evolution. These will be especially problematic if, as seems likely, each galaxy type evolves differently (e.g., Guiderdoni & Rocca-Volmerange 1987; Yoshii & Peterson 1991). The problem is even less constrained if additional galaxy types were visible at earlier cosmic times (Cowie, Songalia & Hu 1991; Cowie *et al* 1995). Observationally, there are possible problems associated with the limiting isophote for image detection, especially for dwarf and intrinsically low surface brightness galaxies (Ferguson & McGaugh 1995; Phillipps & Driver 1995), and with image crowding and overlap (Tyson 1988; Metcalfe *et al.* 1995a).

One way to sidestep many of these difficulties would be to count only galaxies of a single well determined type, whose evolution is thought to be relatively well understood, and which have fairly high surface brightnesses even when seen at high-$z$.

With the tremendously improved imaging quality of the replacement Wide Field and Planetary Camera 2 (WFPC2) on board the Hubble Space Telescope (HST), such a program has now become feasible. The morphological characteristics of faint field galaxies are clearly revealed with WFPC2's $0.1''$ FWHM resolution (Griffiths *et al.* 1994), allowing morphological segregation of field galaxies down to faint magnitudes, while the problem of 'seeing' running neighboring images into one another is all but removed. In particular, elliptical galaxies with high surface brightness and (relatively) simple evolutionary properties perhaps offer the ideal population with which to test and constrain world models. In this paper, we discuss the practicalities of using such a dataset, by applying the techniques to the currently available data, and assess whether the problems inherent in the determination of cosmological parameters can indeed be overcome by this route. In §2 we discuss the currently available data suitable for such an analysis, in §3 we discuss the World models we wish to test, in §4 we discuss how the models are generated and compare these to the observations in §5. Finally we present a discussion and our conclusions in §6.

## 2. The WFPC2 Data for Ellipticals

Three recent HST studies have been made, as part of the Medium Deep Survey (MDS), which present complete magnitude limited samples separated by morphological type and are hence suitable for our purposes. These surveys are: Casertano *et al.* (1995; CRGINOW), Driver, Windhorst & Griffiths (1995; DWG) and Glazebrook *et al.* (1995, GESG). A further ultradeep survey, based on the single deep WFPC2 field of Windhorst & Keel (1995), extends the available data to yet fainter magnitudes (Driver *et al.* 1995;

DWOKGR). The data for the three MDS surveys (summarized in Griffiths *et al.* 1994), constitute randomly selected fields, collected in parallel mode at mostly high galactic latitudes. The CRGINOW data is based on the entire WF/PC database comprising 13,500 galaxies in a magnitude limited sample to $m_{I785LP} = 21$ mag (where the suffix indicates the band and start wavelength of the HST filter). The CRGINOW galaxy images were classified using a Likelihood function to decide whether the measured properties are best matched by a de Vaucouleurs $r^{\frac{1}{4}}$-law, an exponential disk law, or the WF/PC point spread function, resulting in classifications into: bulge-dominated (E/S0), disk-dominated (Sabcd), stellar or unknown (Irr?) systems.

However, for the three fainter WFPC2 surveys, which comprise typically 150-300 galaxies each, the morphological classifications are made by eye. In these surveys, the *V* and *I*-band greyscale plots, in conjunction with major axis light-profiles, are classified by a number of independent "eyeball" estimations. The final classification is then a consensus of these independent classifications. A full spectroscopic follow-up is currently underway to help verify the accuracy of the classifications.

The three WFPC2 surveys classify the faint population into three broad classes: Ellipticals and S0s (E/S0), Early-type Spirals (Sabc) and Late-types/Irregulars (Sd/Irr). The DWG and GESG surveys both span the magnitude range $18 < m_{I814W} < 22$ mag, while DWOKGR, extend the morphological classifications to $m_{I814W} < 24.25$ mag. The DWG and DWOKGR samples make no attempt to distinguish between stars and compact ellipticals, and leave this for the spectroscopic follow-up to determine. This decision was made on the basis of a significant number of non-zero redshifts found for WFPC2 point sources by Mutz *et al* (1994) and Schmidtke *et al.* (1995). To compensate, we choose to subtract the predicted numbers of stars directly, using calculations by Dr. J. B. Jones with software based on a thin plus thick disk and bulge model (Gilmore & Reid 1983), which was kindly supplied by Dr. G. Gilmore. This model should be similar, for our purposes, to the model of Bahcall & Soneria (1980), which include only a thin disk and bulge. The DWG data are adjusted for the sum of the predicted stellar contribution from the six lines-of-sight, and DWOKGR similarly for the single moderately high-galactic latitude field. In all cases, the stellar subtraction results in moving the data points by less than their original Poisson errors.

After stellar subtraction, the agreement between the four independent surveys in the regions of overlap is remarkably good, and the errors in classifications have been assessed at ∼1 Hubble class from comparisons between individual classifiers (c.f. DWOKGR). DWG presented an automated classification technique based on the concentration index of a galaxy's light, which identified the E/S0 population as distinct and separable from the spiral population. DWOKGR also applied the automated methods of CRGINOW, and found good agreement between the eyeball and automated bulge/disk classifications.

For the purposes of constraining cosmological models (and contrary to the usual case for deep galaxy counts), we will be most interested in eliminating models which over-predict the counts. We are thus only interested in how many E/S0 galaxies we may have missed in the classification process. Misclassification between E/S0s and Sd/Irrs is extremely unlikely, given the quality of the HST data. Although a chance alignment may cause an E/S0 to appear Irregular, the statistical likelihood of this is low (based on the isophotal sizes and surface densities of Late-types(Sd/Irr) and Ellipticals(E/S0) from DWG and DWOKGR, we estimate a $\sim 1-2$ % probability of a chance alignment between these types). Errors are thus expected to be restricted to the division between E/S0 and Sabc bins. Furthermore, due to HST's limited SB-sensitivity, it would seem more likely for an Sa galaxy to be misclassified as an E/S0 rather than vice-versa. For example, the disk of a face-on Sa may become lost in the background noise, whereas an elliptical is unlikely to coincide with a positive sky-brightness fluctuation such that a disk is mimicked. Thus we might expect that, if anything, we have already overestimated the real numbers of E/S0s.

Nevertheless, it is prudent to include a consideration of a much more pessimistic classification error. Assuming S0—>Sa as the most probable misclassification, we will assume that the Sa's represent one third of the Sabc galaxies, and propose that the maximum error is derived by assuming that half these Sa's are in reality misclassified E/S0s, and that all objects currently classified as E/S0 are correct.

## 3. The World Models

As originally stated, our main goal is to examine some of the currently popular cosmological models to see if any can be rejected by the current, or foreseeable, data. To model elliptical galaxy counts, we need an input luminosity function (LF), including its local normalization, K-corrections and evolutionary (E-) corrections, and cosmological relations between look-back times, redshifts and volume elements. We choose the elliptical galaxy LF of Marzke $et~al.$ (1994a, b), which is well fitted by a Schechter (1976) function with $M_B^* = -20.5$, $\alpha = -0.9$, $\phi_* = 1.14 \times 10^{-3}$ Mpc$^{-3}$, for $H_0 = 50$ km s$^{-1}$ Mpc$^{-1}$. This agrees quite closely with the shape of the galaxy LF for E/S0s found by Binggelli, Sandage & Tammann (1988) in the Virgo Cluster, and agrees reasonably with the field LF for E/S0s given by Efstathiou, Ellis & Peterson (1988). However, it falls off much less steeply at the faint end than that given by Loveday $et~al.$ (1992) ($M_B^* = -21.2$, $\alpha = +0.1$, $\phi_* = 4.0 \times 10^{-4}$ Mpc$^{-3}$). This is most likely due to incompleteness in the Loveday $et~al.$ elliptical sample (J. Loveday 1995, private communication). Since the LF is falling at faint $M$, the counts for elliptical galaxies are not as critically dependent on the tail of the LF as they are for later type galaxies (see Driver $et~al.$ 1994; Driver & Phillipps 1995; Driver & Windhorst 1995). Nevertheless, there is clearly some uncertainty in the elliptical LF, so we test some models with $\alpha$ at the Loveday *et al.* value in order to assess its importance. We consider the normalization of the LF separately below. We assume throughout that $(B - I) \simeq 2.3$ mag for elliptical galaxies at z = 0 (see the models described in Windhorst *et al.* 1994).

We obtain K-corrections by directly convolving the total HST/WFPC2 response function (*i.e.* optics + filter + detector) with a (non-evolved) elliptical galaxy spectrum[3]. Furthermore, all models are calculated making explicit allowance for isophotal sizes and magnitudes at the appropriate detection thresholds. We assume that the radial intensity profiles follow the usual de Vaucouleurs (1948) $r^{1/4}$-law for ellipticals, and take a range of central surface brightnesses appropriate for such objects, allowing for K-corrections and cosmological surface brightness dimming as a function of redshift (see Bristow & Phillipps 1995 for further details of the modelling). Where included, luminosity evolution is assumed to change the surface brightness at fixed size, since we are considering only passive aging of the stellar population. Note that even high surface brightness ellipticals can start to fall below the isophotal threshold at sufficiently high $z$, because of the large K-corrections, which occur once the blue region of the spectrum is shifted to the observed bandpass (c.f. Pence 1976; King & Ellis 1985). This is ameliorated somewhat by the fact that all the HST samples are based on *I*-band selected data, which does not sample the 4000Å break until $z \geq 1$.

For models with luminosity evolution, we take the simple Tinsley & Gunn (1976)

---

[3]Strictly speaking, we need, a combined K- and E- correction corresponding to $F(\lambda/(1+z), t(z))/(1+z)F(\lambda, t_0)$ made up of a combination of the unevolved K-correction, the change in luminosity with time t(z), at fixed $\lambda$, and the evolutionary change in the galaxy's spectral shape. Here we shall neglect this subtlety as the spectral shape changes only slightly at the red end. The convolution which determines the K-correction also confirms the adopted E/S0-galaxy $(B - I)$ color.

prescription, in which all stars form in an initial burst at high $z$, and the subsequent passive evolution of a galaxy's luminosity (at red and near infra-red wavelengths, at least) is dominated by the variation in the number of red giant stars. This leads to evolution with cosmic time of the form:

$$L(t) = L(t_0)[(t - t_f)/(t_0 - t_f)]^\gamma \quad (1)$$

where $t_f$ is the time of elliptical galaxy formation and $\gamma = -1 + \theta x$ (see Phillipps 1993 and references therein), $x$ being the slope of the initial mass function (IMF), i.e., $x = 1.35$ for a Salpeter mass-function, and $\theta$ is the slope of the mass - main sequence lifetime relation ($\theta \simeq 0.26$). Reasonable values for $\gamma$ are then around $-2/3$. Since for observable redshifts up to z=1 at $I \sim 24$ mag, $t \gg t_f$ we can approximate further and write

$$L(t) = L(t_0)(t/t_0)^\gamma \quad (2)$$

Allowing for a smaller $t_f$ will increase the evolutionary corrections slightly, but even if the formation redshift were as low as $z_f = 2 - 5$, this would amount to only 40% higher values at $z \simeq 1$. A more sophisticated model such as that of Bruzual & Charlot (1993) gives similar evolutionary rates, well approximated by a power law of slope $\gamma \simeq -0.8$ (c.f. their Figure 1; see also Arimoto & Yoshii 1987), or about $\gamma \simeq -0.75$ after allowing for the slight difference in evolution between $V$ and $I$ at moderate redshifts. Hence by adopting a high value for $z_{form}$ we are potentially *underestimating* the level of evolution and once again erring on the side of caution.

The evolutionary effect then translates into different variations of $L$ with $z$, depending on the cosmological model chosen. This simple model does not require specific knowledge of $H_0$, since this scales out in the ratio $t/t_0$, although the actual value of $z_f$ does depend on $H_o$ (c.f. Windhorst, Koo & Spinrad 1986). The specific value chosen above (*ie.* h =0.5) does not, therefore, have any significant influence on the counts generated by the models shown here.

For the conventional cosmological models with zero cosmological constant, we choose density parameters $\Omega_0 = 0.05$ (roughly the value of the baryon density required by primordial nucleosynthesis; see the summary in Bristow & Phillipps 1994) representing a low density universe, and $\Omega_0 = 1$, a flat Einstein-de Sitter universe with critical density, compatible with inflation. In general, low $\Omega_0$ universes have larger volumes out to a given (luminosity) distance, and so generate higher galaxy counts. In all our models, the comoving volume of a redshift shell is given by

$$dV \propto D_L^2 (1+z)^{-2} g(z) dz, \qquad (3)$$

where $D_L$ is the luminosity distance and $g(z)$ is the derivative of coordinate distance with respect to redshift (so that $g(z)dz$ is the comoving thickness of the redshift shell), and $g$ decreases with increasing $\Omega_0$.

For $\Omega_0 = 1$, the scale factor $R \propto (1+z)^{-1} \propto t^{2/3}$, so that the evolutionary correction is simply $L \propto (1+z)$. For the low $\Omega_0$ case, a sufficiently accurate approximation is to take the limiting $\Omega_0 = 0$ expression $R \propto (1+z)^{-1} \propto t$, so that $L \propto (1+z)^{2/3}$.

Recently, several authors (Fukugita *et al.* 1990; Yoshii *et al.* 1993; Yoshii 1993) have suggested that the steep observed $B$-band counts might be better fit by models with a non-zero cosmological constant $\Lambda$. A particular case favored on other grounds (e.g. compatibility with standard inflation and also via anthropic desirability c.f. Efstathiou 1995) is the spatially *flat* model with a dominant contribution to the curvature coming from $\Lambda$, i.e. one with $\Omega_0 + \lambda_0 = 1$ (where $\lambda_0 = \Lambda/3H_0^2 c^2$), as discussed in detail by Peebles (1984). Here we consider both the simple extreme case with $\Omega_0 = 0$, $\lambda_0 = 1$, and the set of solutions where $\Omega_0 + \lambda_0 = 1$, which allows for varying levels of mass density (baryonic plus dark matter) as implied by observations of the mass-to-light ratios of galaxy clusters and the cosmic virial theorem (c.f. Peebles 1980 and references therein). Note that $\Lambda$-dominated models such as these have been suggested as solutions to the 'globular cluster age problem' (i.e. globular clusters being apparently older than the universe, c.f. Pierce 1991; Pierce *et al.* 1994; Freedman 1994) and other perceived difficulties with standard Big Bang models (see summary in Phillipps 1994 and also Bachall 1994), so it would be intriguing if such models were found to be incompatible with the deep HST count data as well. It has previously been suggested (eg. Gardner, Cowie & Wainscoat 1993) that the $\Lambda$-dominated models that fit the $B$ counts (because they have large volumes at high $z$) necessarily overpredict the deep $K$-band counts, which are much less steep, but this has not yet been firmly established (Lilly *et al.* 1991; Djorgovski *et al.* 1995).

In the extreme $\lambda_0 = 1$ model, we get:

$$D_L = (c/H_0)z(1+z) \qquad \text{and} \qquad (4)$$

$$g = c/H_0 \qquad \text{(cf. Phillipps 1994),} \qquad (5)$$

so the volume element reduces to the pseudo-Euclidean case of $dV \propto z^2 dz$. For the more general flat, non-zero $\Lambda$ model we get:

$$D_L = (c/H_0)(1+z)\int_0^z (\Omega_0(1+z')^3 + \lambda_0)^{-1/2} dz' \qquad \text{and} \qquad (6)$$

$$g = (c/H_0)(\Omega_0(1+z)^3 + \lambda_0)^{-1/2} \qquad \text{(eg. Paczynski \& Gorski 1981)} \qquad (7)$$

This latter integral has no simple analytic form (see Charlton & Turner 1987), and must therefore be evaluated numerically.

In summary, then, we have a range of possible cosmological models, either with no spectral-evolution or with passive evolution following the simple Gunn-Tinsley law, and with either a LF slope of $\alpha = -0.9$, or a much more rapidly declining slope of $\alpha = +0.1$. In the following sections, we compare these with the current deep HST data for E/S0s, and assess their differences with regard to forseeable data of similar type.

## 4. Comparing the Models to the HST Data

With the deep HST data and models in place, one final hurdle remains before direct comparisons can be made, namely at which magnitude to normalize the galaxy count models to the data. A now long standing and increasingly worrisome problem is the apparent local underdensity of galaxies revealed by the local redshift surveys (c.f. Loveday *et al.* 1992), which cause all models, irrespective of their cosmological basis, to *under*predict the observed galaxy number-counts at $m_b \simeq 18$ mag. Whether we live in a local hole as proposed by Shanks *et al.* (1990), whether the local data are plagued by systematic errors or incompleteness (Metcalfe *et al.* 1995b; Ferguson & McGaugh 1995), and/or whether significant levels of evolution have occurred locally (Maddox *et al.* 1990a) is unknown. The stellar population studies, at least of nearby ellipticals, suggest little evidence for significant recent luminosity evolution, as does the almost constant mean $(b_J - K)$ color over bright magnitudes (Yoshii & Peterson 1995). Here we follow convention, and renormalize all our proposed models at $m_b \simeq 18$ mag. At this magnitude we are sampling a large enough volume to be approximately homogeneous, but not such a great look back time as to anticipate major evolution. Note though, that all the possible explanations described above are luminosity and/or type dependent, suggesting that a uniform re-normalization is explicitly naive. See DWG for a more detailed discussion of this LF normalization problem.

The models are first normalized to the B-band, as these surveys are both more numerous and extensive at bright magnitudes (Maddox *et al.* 1990a,b; Shanks *et al.* 1984). To normalize each model then, we initially generate a "simulated Universe" equivalent to a 200 square degree wedge, and "observe" it through the detector plus filter response function equivalent to the photographic b-band ($b_J$). The required increase in the LF normalization ($\phi_*$) is then empirically derived by comparing the predicted model counts at $b_J = 18$ mag to the observed value of 8 galaxies per deg$^2$ (in the flat interval $17.75 < b_J < 18.25$ mag).

This normalization value is taken from the average of the APM survey (Maddox *et al.* 1990a) and the CFA1 & 2 catalogs (Marzke *et al.* 1994b) multiplied by $0.40 \pm 0.03$, the local fraction of field E/SOs derived from the Zwicky catalog (Shanks *et al.* 1984) and the CFA1 and 2 surveys (Marzke *et al.* 1994a,b).

The final models are then re-generated for a 1 square degree wedge using the new normalizations, but are now "observed" through the HSTs I814W total response function. To construct the I-band LF, the Schechter function values quoted in the previous section are used except that the $M_*$ value is shifted by 2.3 mags (to correct for the (B-I) color at z =0, c.f. Windhorst *et al.* 1994) and the normalization increased as outlined above. Table 1 summarizes the models and the required renormalizations.

## 5. Models and Results

### 5.1. $\Lambda = 0$ Models

Consider first the standard cosmological models with no evolution (Figure 1). As is well known, decreasing the density parameter from $\Omega_0 = 1$ to $\Omega_0 = 0.05$ steepens the predicted galaxy counts. Different models can have different effective survey depths at a given magnitude. For our models the change from low to high density results in a 50% increase in the counts at $I = 23.75$ mag, our faintest reliable point. The data points follow the $\Omega_0 = 1$ prediction (solid line) quite closely all the way from $I = 18.5$ to $I \simeq 24$ mag, though at the low end of the allowable range of counts. Error bars here are purely from Poisson counting statistics, given the size of our survey areas. A better overall fit would perhaps be achieved if we allowed a slightly higher normalization (by $\sim 20\%$) than given in Table 1 (although the fit at $b_J \sim 18$ mag would be compromised).

The $\Omega_0 = 0.05$ line (short dashes) is also a good fit to the data (especially if we allow a slight *downward* renormalization by $\sim 15\%$, thus moving closer to the local normalization – see Table 1). We note that the galaxy count slopes predicted in our models are slightly flatter than would be the case if we had ignored isophotal effects (i.e., "detected" objects on the basis of their total magnitude rather than their isophotal magnitude). Nevertheless, the slope of the counts at the faint end do differ substantially between the $\Omega_0 = 1$ and the $\Omega_0 = 0.05$ models, so accurate data extending 2 magnitudes could in principle discriminate between them (especially with larger survey areas to reduce counting errors). Other possible ways of discriminating between these two models are discussed in §6.

### 5.2. Spectral Evolution

Let us now consider the situation when we include passive stellar evolution for E/S0s. Strictly this should be considered as the base model, since passive aging of the stellar populations certainly occurs, in addition to whatever else may be going on (see e.g., Franceschini *et al.* 1994). For both the $\Omega_0 = 1$ and $\Omega_0 = 0.05$ models, this passive evolution (luminosity varying as $(1 + z)$ or $(1 + z)^{2/3}$, respectively) has a small but noticeable effect (c.f. thick and thin lines on Figure 1). In fact, in both cases, it equates to a vertical renormalization of about 30% (or 20%) beyond $I \simeq 19$ mag. Allowing for the actual uncertainty in the LF normalization adopted, the $\Omega_0 = 1$ model with passive evolution, perhaps the least controversial of all possible models, fits the HST data remarkably well throughout the whole range $I = 18 - 24$ mag. Correspondingly, the evolving $\Omega_0 = 0.05$ model now significantly overpredicts the observed E/S0 counts, although remains just consistent with the maximum possible HST galaxy classification errors. A downward rescaling by a factor $\sim 1.5$ would now be needed to fit the faintest data points (assuming the HST classifications are indeed correct). Deeper HST images and more precise classifications might definitely rule out this (otherwise attractive) open model. One way to 'save' it might be to adopt a steeply falling LF at the faint end, as in Loveday *et al.* (discussed in §5.5).

## 5.3. Mergers

Unfortunately, stellar evolution might not be the whole story as regards elliptical galaxy evolution. As has been widely discussed, numerical simulations of galaxy mergers convincingly generate elliptical looking galaxies (eg. Barnes 1992), while a significant fraction of ellipticals show evidence for previous merging events (Schweizer & Seitzer 1992). What we need, though, is the overall effect on the brightness of ellipticals in general. If ellipticals steadily accrete small neighbors, then their total mass might be expected to grow linearly with time. If the initial mass was small (i.e. sudden growth from a large number of more or less equal sub-units), then we would have roughly $L(t) \propto t$, while if the accretion was onto an existing large mass, then $L(t) \propto t + const.$, i.e. a slower rate of fractional increase. Similarly, if most of the accretion took place at early cosmic times, the recent (ie. observable) growth would again be slower than $t$.

Kauffmann & White (1993) have made detailed theoretical predictions of the hierarchical growth of massive galaxies (or more precisely their halos). Their massive ($M_*$) merger products had a progenitor of 0.3 to 0.9 of the final mass by $z \simeq 1$ ($t/t_0 \simeq 0.4$), growing to 0.6 to 1.0 times the final mass by $z \simeq 0.5$ ($t/t_0 \simeq 0.6$). Similarly, Lacey & Cole (1993) have shown that present day large galaxies should have reached half their present mass by $t/t_0 \simeq 0.3$. If we now assume that merger-induced evolution scales as $L(t)/L_0 = (t/t_0)^\beta$, then these models would suggest that $\beta \lesssim 1$, with $\beta \sim 0.5$ representing the best estimate from hierarchical clustering (c.f. Kauffmann & White 1993). This would tie in with the amount of merging suggested observationally by Burkey *et al.* (1994) from HST data on close pairs of galaxies at $z \simeq 0.5$.

For the standard cosmological models, such merger evolution effectively counteracts the stellar evolution almost exactly. To see why, let us consider Burkey *et al.* (1994) and Carlberg, Pritchett & Infante (1994), who both find similar epoch dependant merger rates $\propto (1+z)^{3.5}$ or $\propto (1+z)^{2.7-2.4}$ in the co-moving rest frame[4]. Burkey *et al.* find 34 % of the field galaxies to have close companions at redshifts z 0.5 — 0.7. Assuming the most drastic case, this would imply that the total luminosity output from the sum of ellipticals in the co-moving redshift shell cannot be reduced by more than 66 % (*i.e.* this assumes that all close pairs are imminent mergers and in their pre-merger state contribute zero luminosity to the elliptical galaxy counts). Adopting the parameterization above then implies $\beta = 0.8$, which almost but not quite, cancels out the level of evolution expected for purely passive spectral evolution. Thus the model lines move back to (or at least towards) the original no-evolution plots in Figure 1, and our conclusions are unchanged from those originally presented.

### 5.4. $\Lambda \neq 0$ Models

Figure 2 shows the $\Lambda \neq 0$ models for $\lambda_0 = 1$, $\lambda_0 = 0.8$, $\lambda_0 = 0.5$ and $\lambda_0 = 0.2$ with no-evolution (or as we saw above equivalent to an evolution in which the smaller luminosity at higher z due to mergers in progress closely balances the luminosity gain from passive stellar evolution). Fig. 2 shows us that the $\lambda_0 = 1$ model, grossly overpredicts the counts, with a substantial excess even at $I = 21$ mag. By $I = 24$ mag, the $\lambda_0 = 1$ model predicts 2.5 times as many elliptical/S0 galaxies as are seen. In general, the slope is far too steep at

---

[4]Note, however, that the more recent study by Woods *et al.* (1995) finds no evidence for close companions above a randomly distributed sample. While no doubt these observations will be refined it seems prudent here to assume the maximum observed level of galaxy merging

the faint end, confirming the conclusion of Gardner *et al.* (1993) from $K$-band counts.

Taking the more realistic $\lambda_0 = 0.8$ case (as favored recently by Yoshii & Peterson 1995 – also based on $K$-band data), the predicted counts are significantly reduced, but are still much too high and steep (in particular, they are still higher and steeper than in the conventional low $\Omega_0$ model). However, as the model is 'just' consistent within our final HST classification uncertainties, we shall adopt $\lambda_0 = 0.8$ as our *absolute* upper limit. From Figure 2 then we see that the most consistent fits to the data points come from the range of models with $\lambda_0 < 0.2$. Of course a low-$\Lambda$ may be largely defeating the object of the exercise, as the age of the universe is little higher in these models than in conventional ones (c.f. Figure 3).

In fact it is worth digressing to consider the relationship between the age of the universe and $\lambda_0$ in a flat (i.e., k=0) universe. From Carroll, Press & Turner (1992), and adopting k=0, we get:

$$t_0 \approx \frac{2}{3} H_o^{-1} \frac{sinh^{-1}[\sqrt{(1-\Omega_0)/\Omega_0}]}{\sqrt{(1-\Omega_0)}} \qquad (8)$$

which of course is exact for the case $\Omega_0 = 1$. Figure 3 shows this relationship plotted in terms of $\lambda_0$ (*i.e.*, $\equiv 1 - \Omega_0$ for $h = 0.8$, Freedman *et al.* 1995). Our firm absolute upper limit of $\lambda_0 = 0.8$ yields 13.3 Gyrs for the age of the Universe. Taking our optimal fit of $\lambda_0 \leq 0.2$, defined by the Poisson errors alone, gives $t_{age} \leq 9$ Gyrs. Hence, with a value for $H_o = 80$ kms$-1$ Mpc$^{-1}$ (c.f. Pierce 1994; Freedman 1994), our most generous estimate of $\lambda \leq 0.8$ does not solve the recent problem of the Age of the Universe being apparently less than the implied ages of the oldest Globular clusters (summarized for example in van den Bergh 1994).

## 5.5. Modelling uncertainties in $\Lambda = 0$ and $\Lambda \neq 0$ models

Before we can absolutely rule out any models, let us first consider uncertainties in the modelling process. As hinted at above, reducing $\alpha$, the faint end slope of the field LF for ellipticals and S0s, will of course reduce the predicted number-counts at faint magnitudes. Figure 4 shows both the $\lambda_0 = 0$, $\Omega_0 = 0.05$ and the $\lambda_0 = 0.8$, $\Omega_0 = 0.2$ models with $\alpha = +0.1$ (*i.e.* that derived by Loveday *et al*). From Figure 4, we see that for the $\Lambda = 0$ low-$\Omega$ model this change in $\alpha$ is more than enough to reconcile the model with the HST data, showing a strong modelling dependency on the input LF. However for the $\lambda_0 = 0.8$ model still lies well outside the Poisson errors, showing that despite this dependency, the high-$\Lambda$ models still fail to give a close fit to the data points.

Ultimately, to force the high $\lambda_0$ models to become consistent with the HST data, yet more drastic (anti)-evolution is required, such as extremely strong merging (recall that already we are assuming merging to negate passive evolution which is greater than that observed). The previously adopted merging model was such that the decrease in luminosity from fragmenting (as we look back) directly cancels the luminosity enhancement from passive stellar evolution. If we now adopt a much more severe merging model, high-$\Lambda$ models may still remain permissable. The dashed line on Figure 4 shows such an accelerated merging model for a flat $\lambda_0 = 0.8$ universe, with a *combined* net luminosity evolution going as $(1+z)^{-1}$ (*i.e.* passive luminosity evolution $L_{P.E.}(t) \propto (1+z)^1$ and merger-driver luminosity evolution $L_{M.E.}(t) \propto (1+z)^{-2}$ !). That is, fragmenting much more than makes up for the increase in stellar luminosity as we look back. Roughly speaking, stellar evolution would have made galaxies nearly twice as bright at $z = 1$, so we would require each one to have been only a quarter of its present mass, a much more dramatic effect than expected either theoretically or observationally (c.f. Kauffman & White 1993; Burkey *et al.* 1994 respectively). Even then, the reduced number count slope is only able to fit the data out to $I = 22$ mag at most (see short dashed line in Figure 4). Very drastic merger evolution indeed would be required to enable these models to fit the counts at our faintest magnitudes. Combining strong merging with a steeply falling LF slope can of course force the models to fit, but is clearly contrived, as they require the very sharply declining Loveday *et al.* LF with $\alpha = +0.1$ (which is almost certainly incomplete at the faint end) *combined with a large merger rate.* (c.f. large dashed line in Figure 4).

## 6. Conclusions

Given the expected levels of stellar and dynamical evolution in elliptical galaxies, our model predictions should lie close to, or perhaps slightly above, our "no evolution" lines shown in Figure 1. As such, the conventional low density ($\Omega_0 = 0.05$) and flat ($\Omega_0 = 1$) models bracket the data points almost perfectly. When purely stellar evolution is included (thick lines in Figure 1), then the $\Omega_0 = 1$ model is preferred. However, the difference between the curves for $\Omega_0 = 0.05$ with no (combined stellar and dynamical) evolution and for $\Omega_0 = 1$ with passive stellar evolution are very small (about the size of our individual $1\sigma$ error bars) even at $I = 24$ mag. Thus, even if we had much larger HST data sets, extending significantly ($\geq 2$ mag) fainter and for which we had accurate classifications, the uncertainties in our theoretical understanding of evolution would still prevent us from discriminating between open and flat standard models. In addition, differences in the model predictions based on different estimates of the local E/S0 LF also appear at the same level, so even if consensus were reached on the precise spectral and luminosity evolution, a low $\Omega_0$ model with a more rapidly declining faint end to the LF can still mimic the $\Omega_0 = 1$ model (c.f. solid line of Figure 4). This reinforces the *dire need* for accurate *local* LF data before we can make far reaching cosmological deductions. In particular, it seems crucial to understand the physical cause of the re-normalization problem discussed earlier. Ultimately then, it seems that E/S0 number-count data alone are insufficient to determine $\Omega_0$ in a $\Lambda = 0$ universe.

The situation with regard to the $\Lambda$ dominated models is more promising, in that even with no (net) evolution, these models significantly overpredict the observed counts for E/S0s; with net positive evolution (*i.e.* galaxies brighter in the past), this situation would of course be worse. In order to reconcile the $\lambda_0 = 0.8$, $\Omega_0 = 0.2$ model with the current observations, we would need *both* a drastic evolution in the merger rate, *and* an extreme LF with virtually no faint early-type galaxies (c.f. Figure 4). These may, of course, be mutually incompatible, since the pre-merger objects should have been largely at the faint end of the LF (though one might argue that they were not then ellipticals). The more likely direction of classification errors only makes the situation worse, and we would have to appeal to the extreme (and unlikely) case discussed in section 2 (whereby 1/2 of the galaxies classified as Sa are really E/S0s, with no compensating errors in the other direction) in order to get the no evolution $\lambda_0 = 0.8$ model within the error bars (c.f. Figure 2). The slope of this prediction is still steep, though, so deeper data (or confirmed classifications) should easily rule out some of these models. Even the "half-and-half" model, (*i.e.* $\lambda_0 = 0.5$, $\Omega_0 = 0.5$), predicts many more galaxies than observed at the faintest magnitudes (c.f. Figure 2).

One way to augment these studies would be to include color information, see for example Cowie et al (1994), where they use the number of extremely red objects at faint magnitudes to constrain the density of unevolved objects at high redshifts. A detailed color survey of the HST morphological data would therefore provide additional constraints on the degree of evolution which may be taking place. As yet color information is not available for the existing samples but will be the topic of a future paper by Odewahn *et al.* (1995).

Finally, given the full spectroscopic survey now in progress, it is worth considering the extra constraints that this may provide in terms of redshifts (cf. the methods of Loh

– 21 –& Spillar 1986; Yoshii & Takahara 1989), besides the conformation of the morphological classifications. In particular, do the evolving $\Omega_0 = 1$ and the non-evolving $\Omega_0 = 0.05$ models and the negatively evolving $\lambda_0 = 0.8$ model with a falling LF – which all fit the counts – make different predictions for $n(z)$ ? It is easy to see this theoretically, since (luminosity) distances are smaller at given $z$ in the $\Omega_0 = 1$ model than for $\Omega_0 = 0.05$, so that conversely at given $D_L$ (*i.e.* distance modulus), redshifts are greater in the former case. Including luminosity evolution, too, means that galaxies should be seen at significantly larger $z$ in this model. This is confirmed in the numerical simulations which show that for galaxies with $22 \leq I \leq 24$, $<z> = 0.76$ for $\Omega_0 = 1$ but $<z> = 0.65$ for $\Omega_0 = 0.05$. Perhaps more useful, a substantial number of these faint objects ($\sim 21$ %) should be at $z > 1$ in the flat model, but virtually none ($\sim 4$ %) for the open model (for $22 < I < 24$ mag).

For high $\lambda_0$, distances are even larger than for the open model and the negative evolution required further reduces the redshifts we reach. However, because we also need a sharply declining faint end of the LF to make this model fit, we have very few low-$z$ intrinsically faint objects. The redshift distribution therefore becomes very highly peaked in this model compared to the others. Overall, we conclude that it should indeed be possible to place further constraints and break the degeneracy of the counts alone if and when sufficient redshift data at these faint magnitudes becomes available.

To summarize then, even with perfect data (going say 2 mag deeper than at present), uncertainties in the evolutionary corrections (primarily in the importance of mergers) and in the local LF prevent us from discriminating between open and flat standard models purely on the basis of the E/S0 counts. The addition of redshift information may be able to overcome this ambiguity. On the other hand, the $\Lambda$ dominated models are already in serious trouble, even with just the current data. Extreme choices of the evolution and the local LF are required in order to 'save' them. Improvements in our understanding of either the evolution or the local LF should enable us to conclusively reject them. The very narrow predicted $n(z)$ in this case might also provide a good discriminant. We therefore concur with earlier conclusions (on the basis of $K$-band observations) that explaining the steep $B$-band counts by invoking non-zero $\Lambda$ is not viable due to the failure to match the counts at other wavelengths.

The direct implication is that for a flat cosmology, $\lambda_0$ is firmly constrained to 0.8 or less, implying an age for the universe no greater than 13.3 Gyrs (for $H_o = 80$ kms$^{-1}$ Mpc$^{-1}$). If we further assume that the current morphological classifications of faint field ellipticals (E/S0s) already represents a firm upper limit, as is most probable considering the likely direction of classification error, then $\lambda_0$ is constrained to less than 0.2, which to all extent and purposes is negligible. The implications for the maximum age for a flat universe is the usual $\approx \frac{2}{3} H_0^{-1}$ value of $< 9$ Gyrs, based on $H_o = 80$ kms$^{-1}$Mpc$^{-1}$. It is clear then, from both our pessimistic and optimistic estimations, that invoking a cosmological constant to preserve a flat universe (and therefore a single inflationary scenario) does not solve the $H_o$/globular cluster age problem.

Our one final word of caution, in what we've believed to be an cautious analysis, is that all the models discussed here have been renormalized at $b_J \sim 18$ mag by typically an unexplained factor of 1.5 (c.f. Table 1). Although the need for this renormalization seems well-established, it is somewhat ironic that perhaps the major obstacle for the interpretation of faint galaxy counts is our limited understanding of the *local* universe, namely the field LF and its absolute normalization.

The shallower data in this analysis came from published papers from the HST Medium Deep Survey. We would therefore like to thank Stefano Casertano, Kavan Ratnatunga, Eric Ostrander, and Richard Griffiths for their help in various stages of the MDS project. We also thank Gerry Gilmore for supplying his Galaxy model software and John Bryn Jones for computing the expected stellar contamination. SP and PDB thank the Royal Society and PPARC, respectively, for financial support. RAW acknowledges support from HST grants GO.5308.01.93A and GO.2684.03.93A.

## Tables

Table 1: Summary of the cosmological models tested.

| Model | SF[1] | $\Omega_0$ | $\lambda_0$ | Evol$^n$ | $\alpha$ |
|---|---|---|---|---|---|
| a | 1.70 | 1.0 | 0.0 | None | $-0.9$ |
| b | 1.64 | 1.0 | 0.0 | $(1+z)^1$ | $-0.9$ |
| c | 1.67 | 0.05 | 0.0 | None | $-0.9$ |
| d | 1.63 | 0.05 | 0.0 | $(1+z)^{2/3}$ | $-0.9$ |
| e | 1.56 | 0.0 | 1.0 | None | $-0.9$ |
| f | 1.30 | 0.2 | 0.8 | None | $-0.9$ |
| g | 1.39 | 0.5 | 0.5 | None | $-0.9$ |
| h | 1.46 | 0.8 | 0.2 | None | $-0.9$ |
| i | 2.01 | 0.05 | 0.0 | None | $+0.1$ |
| j | 1.73 | 0.2 | 0.8 | None | $+0.1$ |
| k | 1.46 | 0.2 | 0.8 | $(1+z)^1$ | $-0.9$ |
| l | 1.91 | 0.2 | 0.8 | $(1+z)^1$ | $+0.1$ |

Notes:

[1]SF refers to the scaling factor or model renormalization, *i.e.* the amount by which the local luminosity function must be scaled up by for the relevant model to match the number-counts at $b_J = 18$ mag.

## Figures

Figure 1: Model predictions compared to observations for $\Lambda = 0$, high ($\Omega = 1$, solid lines) and low ($\Omega = 0.05$ dashed lines) mass universes, with (thin lines) and without (thick lines) luminosity evolution. The data shown is for elliptical and S0 galaxies identified on HST WF/PC and WFPC2 images from the Medium Deep Survey (CRGINOW; DWG for $I \leq 22$ mag) and from a single ultradeep WFPC2 field (DWOKGR for $I \leq 24.25$ mag). The classification uncertainty defines the maximum estimated error in the morphological classifications, as defined in the text.

Figure 2: Model predictions compared to the observations for $\Lambda \neq 0$ flat models. A conservative zero net-evolution is assumed. Data as Fig. 1.

Figure 3: The dependency of the age of the universe on the cosmological constant, assuming $\lambda_0 + \Omega_0 = 1$ and $H_o = 80$ km$^{-1}$Mpc$^{-1}$.

Figure 4: Selected models with non-standard LFs and drastic merger evolution to test modelling robustness and dependencies.

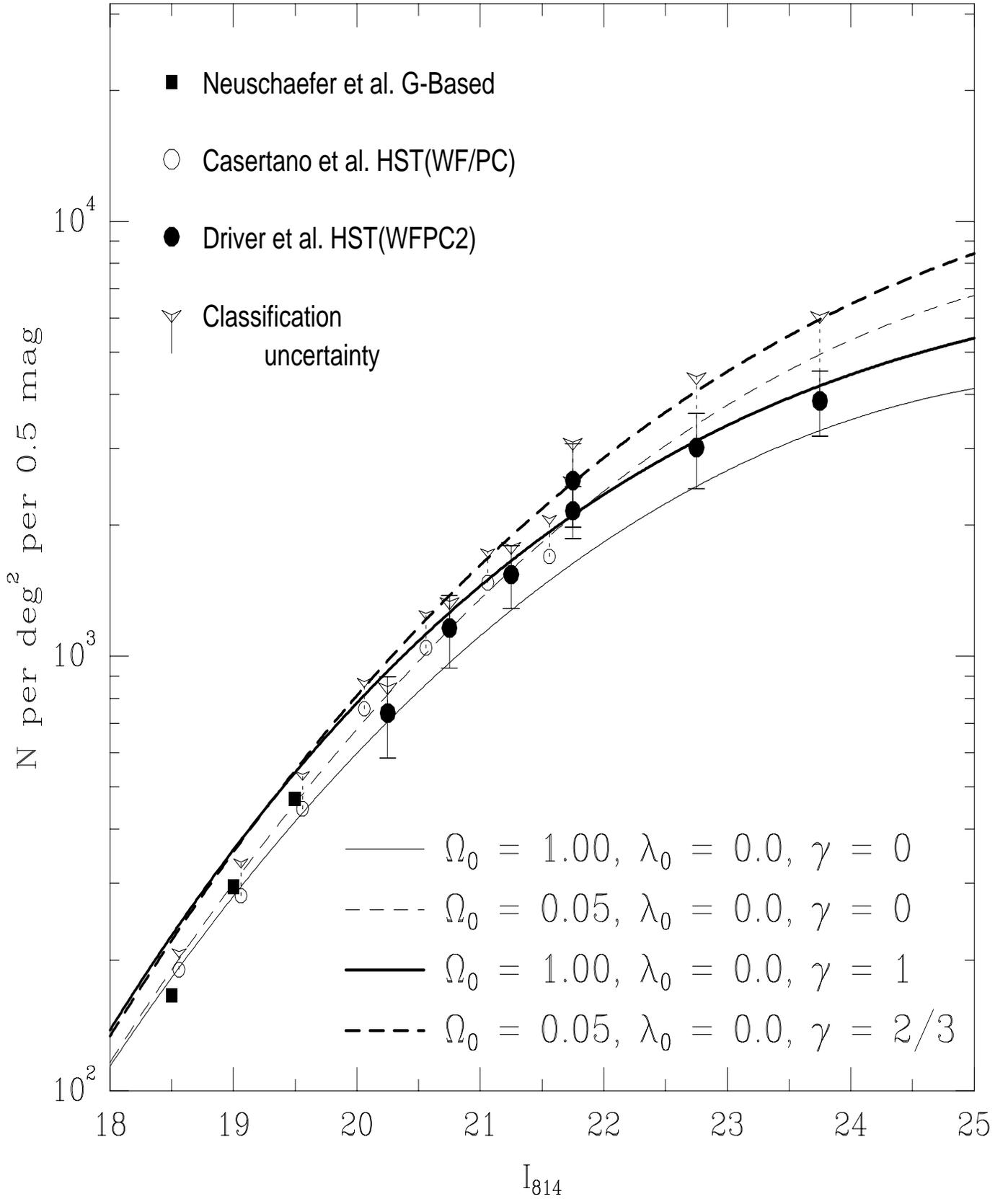

Fig. 1

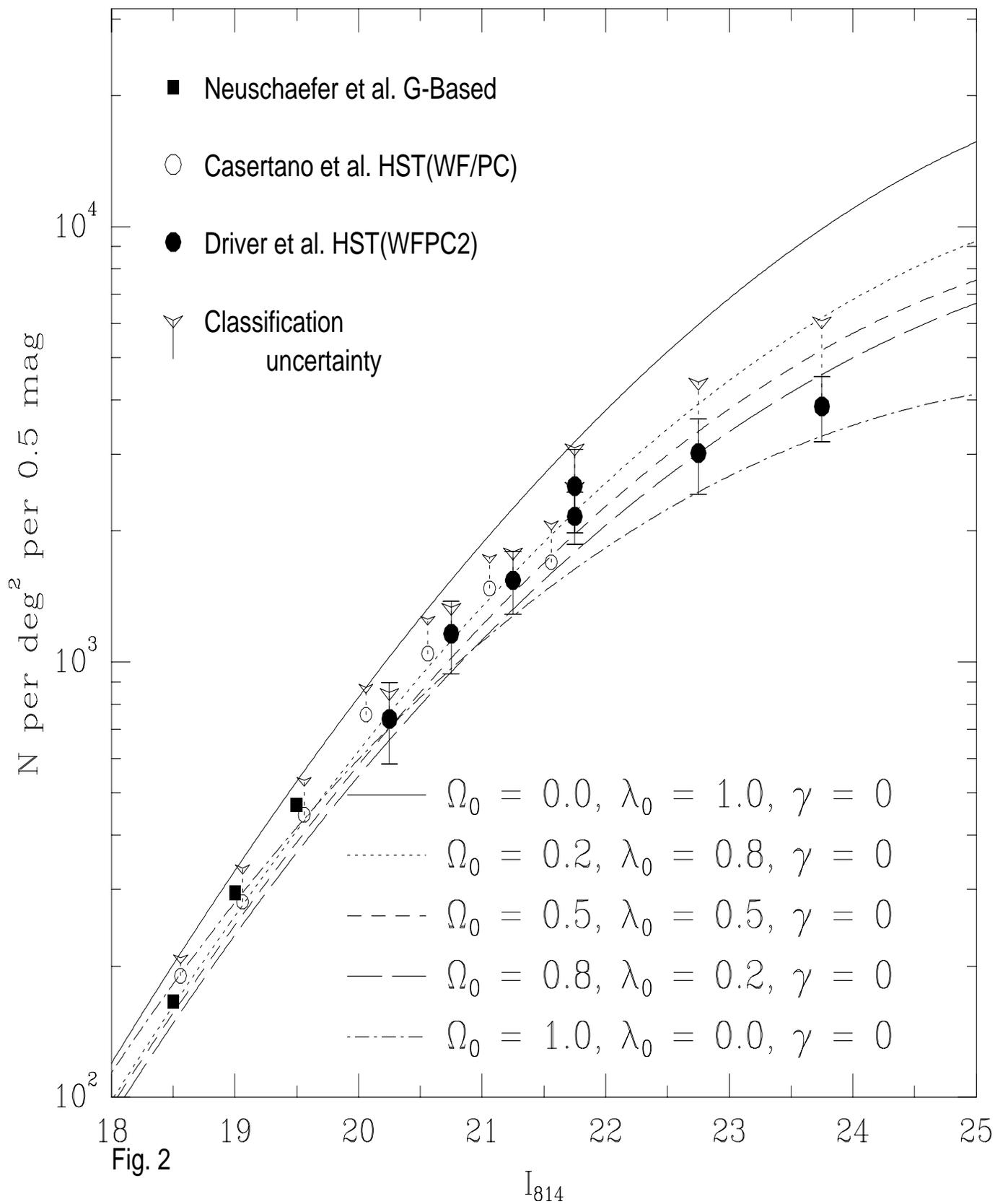

Fig. 2

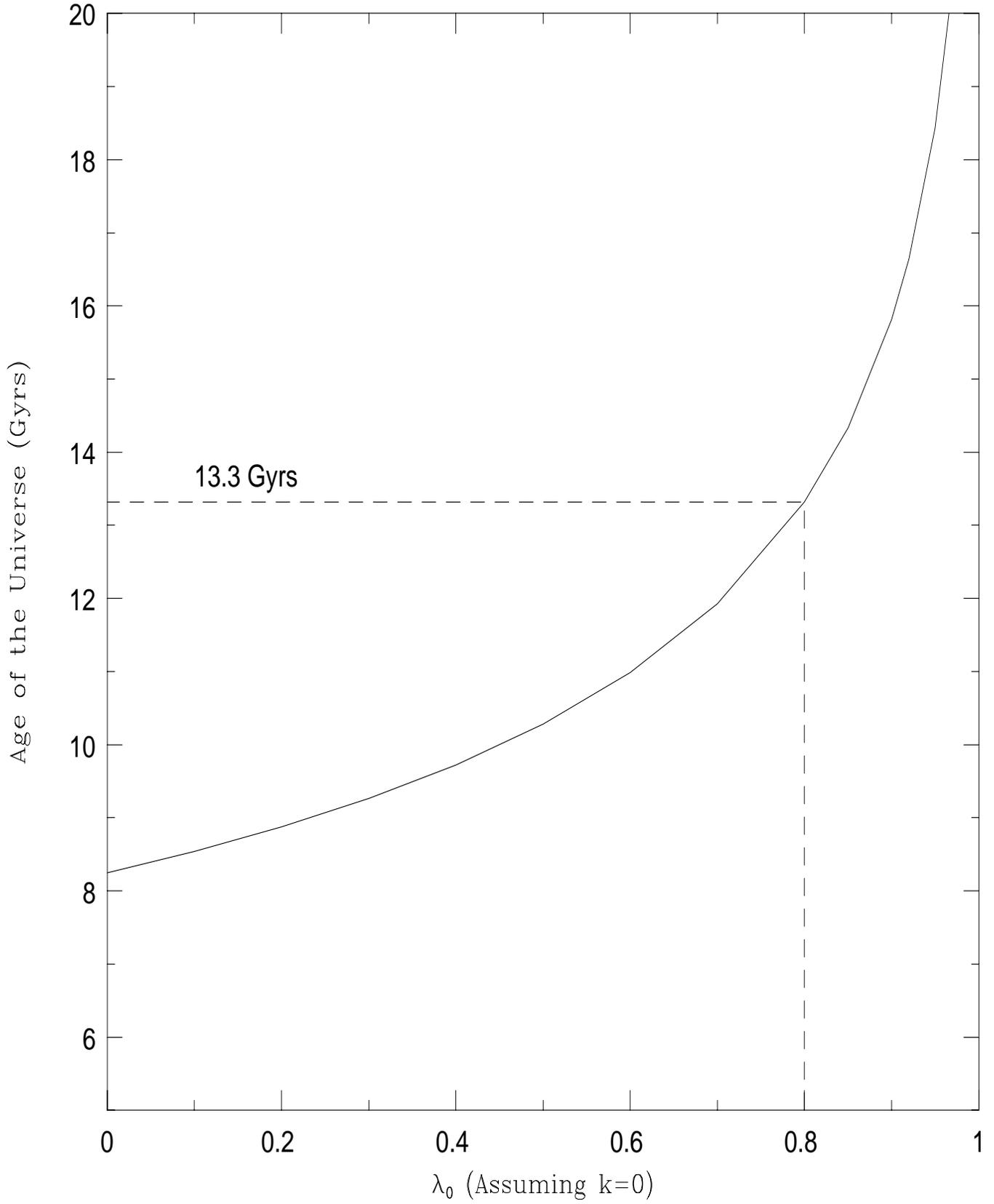

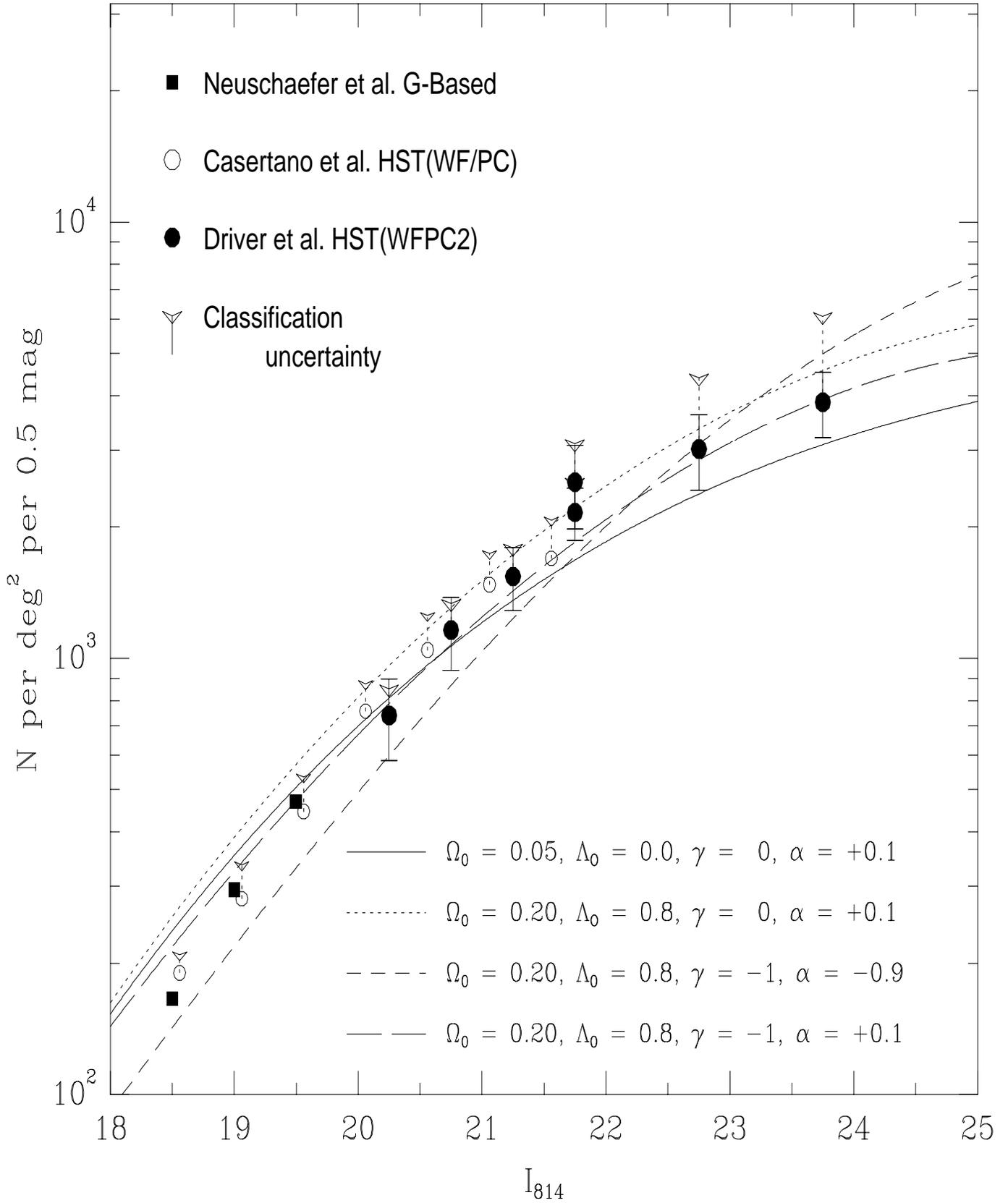

Fig. 4